\documentclass[english,aps,amsmath,amssymb,11pt]{revtex4}
\usepackage[T1]{fontenc}
\usepackage[latin9]{inputenc}
\usepackage{mathrsfs}
\usepackage{amsmath}

\makeatletter
\@ifundefined{textcolor}{}
{%
 \definecolor{BLACK}{gray}{0}
 \definecolor{WHITE}{gray}{1}
 \definecolor{RED}{rgb}{1,0,0}
 \definecolor{GREEN}{rgb}{0,1,0}
 \definecolor{BLUE}{rgb}{0,0,1}
 \definecolor{CYAN}{cmyk}{1,0,0,0}
 \definecolor{MAGENTA}{cmyk}{0,1,0,0}
 \definecolor{YELLOW}{cmyk}{0,0,1,0}
}

\@ifundefined{definecolor}{\usepackage{color}}{}
\usepackage{babel}

\usepackage{latexsym}\usepackage{bm}

\makeatother

\usepackage{babel}
\begin{document}

\title{Holographically Viable Extensions of  Topologically Massive and Minimal Massive Gravity ?}

\author{Emel Altas, Bayram Tekin }

\email{btekin@metu.edu.tr,altas@metu.edu.tr}

\selectlanguage{english}%

\affiliation{Department of Physics,\\
 Middle East Technical University, 06800, Ankara, Turkey}

\date{\today}
\begin{abstract}
Recently \cite{Townsend1}, an extension
of the topologically massive gravity (TMG) in $2+1$ dimensions, dubbed
as minimal massive gravity (MMG), which is free of the bulk-boundary unitarity clash  that inflicts the former theory and all the other known three dimensional theories, was found. Field equations of MMG differ from those of TMG at quadratic terms in the curvature that do not come from the variation of an action depending on the metric alone. Here we show that MMG is a unique theory and there does not exist a deformation of TMG or MMG
at the cubic and quartic order (and beyond) in the curvature that is consistent at the level of the field equations. The only extension of TMG with the desired bulk and boundary properties having a single massive degree of freedom is MMG.

\end{abstract}
\maketitle

\section{INTRODUCTION}
One of the most promising approaches to a quantum theory of gravity  is via the anti-de
Sitter (AdS)- conformal field theory (CFT) \cite{maldacena}
correspondence where there is a boundary field theory dual to the
bulk gravity. In $2+1$ dimensions, where gravity is somewhat less
complicated, this idea has been vigorously pursued in many different works. Einstein's
gravity with a cosmological constant in $2+1$ dimensions is locally
trivial with no propagating degrees of freedom; therefore to study
a dynamical theory which might mimic realistic gravity  and teach us something about four-dimensional quantum gravity, the next option is to consider the parity-noninvariant topologically massive gravity (TMG) which has a single massive
graviton \cite{Deser1}. TMG with a cosmological constant has two copies of Virasoro algebra,
as its asymptotic symmetry algebra, in the two-dimensional boundary
of $AdS_{3}$. In TMG, unitarity of the putative boundary CFT is in
conflict with the unitarity of the bulk theory except, ostensibly, at the chiral
point where the problematic  negative central charge of the boundary field theory
vanishes and the other central charge is positive \cite{strom}.
But, exactly at this point in the parameter space of couplings, there arise 
solutions with asymptotically non-AdS (logarithmic) behavior which cannot be eliminated
from the spectrum on solid physical grounds 
except with ad hoc strong boundary conditions \cite{carlip,grumiller,gribet}. So apparently, TMG
by itself does not have a unitary dual CFT in asymptotically AdS
spacetimes, and hence most probably is not viable as a quantum theory (at
least in the sense of AdS/CFT correspondence).

Another dynamical theory, new massive
gravity (NMG) \cite{nmg},  a judiciously chosen quadratic extension of Einstein's gravity and with two helicity-2 (albeit massive)  
degrees of freedom closer to the four-dimensional gravity, also has the  bulk-boundary unitarity
clash and hence does not posses the expected holographic description. Unfortunately, healthy deformations of NMG in the bulk, such as the
cubic, quartic \cite{sinha} or infinite
order ones \cite{binmg,binmgc,paulos} also suffer from boundary nonunitarity and so probably lack a CFT dual.

With all these negative results, there seems to be an apparent impasse: Einstein's gravity has a healthy boundary structure \cite{brown} but suffers from bulk triviality and all the locally nontrivial theories seem to suffer from bad boundary behavior and so one may wonder if is it not possible to construct
a dynamical theory of gravity in 2+1 dimensions that is unitary
both in the bulk and on the boundary. It turns out that one can actually
construct such a theory \cite{Townsend1} once one gives up the condition that the theory comes from the variation of an action which is purely defined in terms of the metric. (There does exist an action in the first order, that is the dreibein and the spin-connection formulation \cite{Townsend1,Baykal}). Equations
without a proper Lagrangian formulation are not unheard of in macroscopic
physics, but clearly this is a rather novel idea in microscopic phenomena.
But (quantum) gravity is so elusive that one must try many different routes
to get a possible understanding of it. In \cite{Townsend1}
keeping the bulk properties of TMG intact, a theory with an improved
boundary behavior was formulated in terms of consistent field
equations. Namely, the field equations do not have a Bianchi identity
for generic smooth metrics, but they do satisfy a Bianchi identity for
all solutions of the theory. Therefore the theory is consistent as
a classical one and can also be studied as a quantum theory. Its bulk and boundary unitarity and chiral version and conserved charges were constructed in \cite{Alishahiha:2014dma,Tekin:2014jna}. TMG's deformation with two helicity-2 degrees of freedom was constructed in 
\cite{mmg2} called MMG$_{2}$, which also has unitary bulk
and boundary properties for a large class of spacetimes.

The fact that MMG has these remarkable properties which the other three-dimensional theories lack,  begs the question if the theory is unique or if it is part of a large class of theories
that are defined by consistent field equations but do not come from the variation of an action. 
 In this current work, we show several things. First we prove that
at the quadratic order in the curvature MMG is the only possible deformation of TMG.
Our proof will make the rather "magical" appearance of the on-TMG-shell conserved $J$-tensor  more intuitive. Then we move on to the cubic and quartic orders in the curvature and
show in detail that there does not exist a deformation of TMG or MMG
with a single massive degree of freedom.  The Schouten identities  satisfied by the powers of the Ricci tensor guarantee that no new  algebraically independent rank -2  tensors built with the powers of the curvature arise beyond the quadratic terms and  hence the proof is valid for all theories based on the powers of the curvature and not its derivatives. [Note that if derivatives of the curvature are introduced the problem turns into a separate one, diverging from the idea of extending the single massive degree of freedom theories.  MMG$_{2}$ discussed above is an example of that.]

Our construction here basically
answers the following problem. Let ${\cal E}_{\mu\nu}=0$ be the
field equations 
\begin{equation}
{\cal E}_{\mu\nu}=\sigma G_{\mu\nu}+\Lambda_0 g_{\mu\nu}+\frac{1}{\mu}C_{\mu\nu}+\gamma Y_{\mu\nu} =0,
\label{denklem_esas}
\end{equation}
where the Einstein and Cotton tensors read, respectively, as 
\begin{equation}
G_{\mu\nu}=R_{\mu\nu}-\frac{1}{2}g_{\mu\nu}R, \hskip 0.7 cm 
C_{\mu\nu}=\eta_{\mu}\thinspace^{\alpha\beta}\nabla_{\alpha}S_{\beta \nu},
\end{equation}
and $S_{\mu \nu} \equiv  R_{\mu \nu} -\frac{1}{4} g_{\mu \nu}R$  is the  Schouten tensor. The completely antisymmetric tensor is defined  in terms of the completely symmetric symbols as $\eta^{\nu\rho\sigma}\equiv\epsilon^{\nu\rho\sigma}/\sqrt{-\det g}$.
The main question is to find all the possible $Y^{\mu\nu}$ tensors
which satisfy the on-shell conservation: Namely  we demand the on-shell Bianchi identity :
\begin{equation}
\nabla_{\mu}\varepsilon^{\mu\nu}=\gamma\nabla_{\mu}Y^{\mu\nu}=0.      \hskip 0.5 cm       (\mbox{on shell})
\end{equation}
Let us study the problem order by order in the powers of curvature.
\section{ $R^2$-extensions}
Let us assume that  one has  the most general quadratic tensor as 
\begin{equation}
Y^{\mu\nu} \equiv a {\cal{S}}_2^{\mu \nu} +b  g^{\mu\nu}{\cal S}_2 +c  S_{\mu\nu}S +d  g_{\mu\nu}S^{2},
\end{equation}
where, not to clutter the notation, we have defined ${\cal{S}}_2^{\mu \nu} \equiv   S^\mu_\rho S^{\rho \nu}$  and  ${\cal{S}}_2 \equiv   S_{\mu \nu} S^{\mu \nu}$ which will come in handy when more powers of the tensors are constructed. 
The trace and divergence are, respectively,
\begin{equation}
Y = (a + 3 b) {\cal{S}}_2+(c+ 3 d)S^{2},
\end{equation}
\begin{equation}
\begin{split}
\nabla_{\mu}Y^{\mu\nu}=\Big ((a+c)S_{\rho}\thinspace^{\nu}
+(2d+c)S\delta_{\rho}^{\nu}\Big)\nabla^{\rho}S+S^{\mu\rho}\Big (a\nabla_{\mu}S_{\rho}\thinspace^{\nu}+2b\nabla^{\nu}S_{\mu\rho}\Big ).
\end{split}
\end{equation}
For this vector to vanish on the TMG mass shell, we must turn the last part to a Cotton tensor, which is possible only if $a=-2b$
yielding 
\begin{equation}
\nabla_{\mu}Y^{\mu\nu}=\Big ((a+c)S_{\rho}\thinspace^{\nu}+(2d+c)S\delta_{\rho}^{\nu}\Big )\nabla^{\rho}S+a\eta_{\lambda}\thinspace^{\nu\mu}S_\mu\,^\rho C_{\rho}\thinspace^{\lambda},
\label{Ydenk}
\end{equation}and reducing the trace to
\begin{equation}
Y =  b{\cal{S}}_2 +(c+ 3 d)S^{2}.
\label{trace2}
\end{equation}
Here the discussion bifurcates: If $b\ne 0$, then the modified theory does not preserve TMG's property that all solutions have 
$\nabla_\mu R=0$. On the other hand, if $b=0$, the modified theory keeps this property of  TMG intact.  But in the latter case, no new theories arise beyond TMG since one has  
theories are  $Y_{\mu\nu} = c S_{\mu\nu}S +d  g_{\mu\nu}S^{2} $, which for constant $S$ simply gives a shift of TMG parameters. So we assume $b \ne 0$. In this case, in (\ref{Ydenk}), the term with the Cotton tensor vanishes onshell  (\ref{denklem_esas}). The  first term in (\ref{Ydenk}) does not vanish unless one sets 
\begin{equation}
a+c=0, \hskip 1 cm 2d+c=0,
\end{equation}
which reduces, after fixing the overall coefficient as choose $a=-1$,  the $Y_{\mu \nu}$-tensor to the $J$-tensor found in \cite{Townsend1} as  
\begin{equation}
J^{\mu\nu}=-{\cal{S}}_2^{\mu \nu} +\frac{1}{2}  g^{\mu\nu}{\cal S}_2 +S^{\mu\nu}S-\frac{1}{2}g^{\mu\nu}S^{2},
\end{equation}
which can be recast as 
$J^{\mu\nu}\equiv \frac{1}{2}\eta^{\mu\rho\sigma}\eta^{\nu\tau\eta}S_{\rho\tau}S_{\sigma\eta}$ which has the following interesting properties. Its trace is  given as
\begin{equation}
J = \frac{1}{2} \Big ({\cal S}_2- S^2   \Big ),
\end{equation}
which is nothing but the the quadratic part of NMG, the theory that defines a massive spin-2 particle with two helicities.  Quite remarkably,  as noted in \cite{mmg2},  the variation of the quadratic part of NMG splits into two parts as 
\begin{equation}
\delta_g \int \sqrt{-g}\, d^3 x \, J  \equiv  J_{ \mu \nu} + H_{\mu \nu},
\label{action}
\end{equation}
where  the $H$-tensor is 
\begin{equation}
H_{ \mu \nu} \equiv  \frac{1}{2}\eta_{\mu}\,^{ \alpha\beta}\nabla_{\alpha} C_{\beta \nu} + \frac{1}{2}\eta_{\nu}\,^{ \alpha\beta}\nabla_{\alpha} C_{\beta \mu}.
\end{equation}
Clearly one has  $ \nabla_\mu H^{\mu \nu} = -  \nabla_\mu J^{\mu \nu} =  \eta^{\nu\alpha\beta}S_{\alpha \sigma} C_{\beta}\,^\sigma $, and so it follows from (\ref{action}) that $J$- and $H$-tensors  are not separately  automatically covariantly  conserved. But when the  $J$-tensor is augmented to TMG equations, one gets a consistent, on-shell, conservation, which can also be coupled to matter consistently, albeit in a rather complicated way \cite{Arv}. 
We would like to note the following observation:  The $H$-tensor, when looked at with closer scrutiny, is nothing but the  three-dimensional version of the Bach tensor $B_{\mu \nu}$ that measures whether the spacetime is a conformally Einstein manifold or not in four dimensions.  Namely, in $n$ dimensions, the Bach tensor is given as  
\begin{equation}
B_{\mu \nu} =  \nabla ^\alpha \nabla^\beta W_{ \mu \alpha \nu \beta } + \frac{1}{2} R^{ \alpha \beta } W_{\mu \alpha \nu \beta } , 
\end{equation} 
which in this form does not allow a three-dimensional analog since the Weyl-tensor ($W_{ \mu \alpha \nu \beta }$)  vanishes identically. But  an equivalent form of the Bach tensor is 
 \begin{equation}
B_{\mu \nu} = \frac{1}{2} \nabla^\alpha C_{\alpha \mu \nu } + \frac{1}{2} R_{ \alpha \beta } W_\mu\,^{\alpha}\,_\nu\,^\beta , 
\end{equation} 
where $ C_{\alpha \mu \nu }$  is the three index Cotton tensor that serves as a "potential" to the Weyl-tensor and  is defined in any dimension as 
\begin{equation}
C_{\alpha \mu \nu  } = \nabla_{\alpha} R_{\mu \nu} - \nabla_{\mu} R_{\alpha \nu} - \frac{1}{ 2( n-1)} \Big ( g_{ \mu \nu} \nabla_{\alpha } R-  g_{ \alpha \nu} \nabla_{\mu } R \Big ).
\label{3indexcot}
\end{equation}
In three dimensions, one has 
\begin{equation}
C_{\mu \nu} = \frac{1}{2} \eta_{\mu}\,^{ \alpha \beta} C_{ \alpha \beta \nu},   
\label{2indexcot}
\end{equation}
and hence follows the equivalence of  the $H$- tensor and the three-dimensional version of the Bach tensor. This is a rather unexpected result which says that when restricted to the conformally Einstein (or conformally flat, which are the same in three dimensions ) metrics, the quadratic part of NMG reduces to that of MMG. The quadratic part of NMG, without the Einstein term, was studied in \cite{deser_prl} as a separate model.  
Note that in four dimensions the Bach tensor is divergence free but not so in other dimensions, including three dimensions. 

Before we move on to the higher powers, let us give a rederivation of the uniqueness of  MMG .

\section{UNIQUENESS OF MMG}

Suppose  $X_{\mu\nu}$ is a symmetric and divergence-free ($\nabla_{\mu} X^{\mu\nu}=0$) tensor, coming from the variation of an action purely based on the metric.  For the $X$-tensor to be divergence free, the action has to be diffeomorphism invariant at least up to a boundary term as in the case of TMG. We shall denote the traces without  and index as  $X\equiv X_{\mu\nu}g^{\mu\nu}$. Using this tensor, we can build a symmetric two-tensor quadratic    in our given tensor as  
\begin{equation}
Y^{\mu\nu}\equiv\frac{1}{2}\,\eta^{\mu\rho\sigma}\eta^{\nu\tau\eta}\widetilde{X}_{\rho\tau}\widetilde{X}_{\sigma\eta}\,,
\label{Y_tensor}
\end{equation}
where  $\widetilde{X}_{\sigma\eta}=X_{\sigma\eta}+ a\  g_{\sigma\eta}X$ with $a$ real number for now. Note that 
with just one single parameter, the above  $Y$-tensor is in a specific form: The most general quadratic form  reads
\begin{equation}
Y_{\mu\nu}\equiv X_{\mu}^{\rho}X_{\rho\nu}+c_1 g_{\mu\nu}X_{\rho\sigma}X^{\rho\sigma}+c_2 X_{\mu\nu}X +c_3 g_{\mu\nu}X^{2}.
\label{J-tensor}
\end{equation}
But  taking this second form simply  extends the length  of the following computations eventually resulting to the same conclusion. Using 
\begin{equation}
\eta^{\mu\sigma\rho}\eta_{\nu\alpha\beta}=-\delta^{\mu}{_{\nu}}\Big(\delta^{\sigma}{_{\alpha}}\delta^{\rho}{_{\beta}}-\delta^{\sigma}{_{\beta}}\delta^{\rho}{_{\alpha}}\Big)+\delta^{\mu}{_{\alpha}}\Big(\delta^{\sigma}{_{\nu}}\delta^{\rho}{_{\beta}}-\delta^{\sigma}{_{\beta}}\delta^{\rho}{_{\nu}}\Big)-\delta^{\mu}{_{\beta}}\Big(\delta^{\sigma}{_{\nu}}\delta^{\rho}{_{\alpha}}-\delta^{\sigma}{_{\alpha}}\delta^{\rho}{_{\nu}}\Big),\label{identity}
\end{equation}
it is easy to show that  (\ref{Y_tensor}) has all the required tensor structures in it. Hence, we shall start with it.
Then one has $\widetilde{X}=(1+3a)X$, and the divergence of $\widetilde{X}^{\sigma\eta}$ reads
\begin{equation}
\nabla_{\sigma}\widetilde{X}^{\sigma\eta}=\frac{a}{(1+3a)}\nabla^{\eta}\widetilde{X}.
\end{equation}
Using this, one can compute the divergence of $Y^{\mu\nu}$ as
\begin{equation}
\nabla_{\mu}Y^{\mu\nu} = \,\eta^{\mu\rho\sigma}\eta^{\nu\tau\eta}\widetilde{X}_{\rho\tau}\nabla_{\mu}\widetilde{X}_{\sigma\eta}
\equiv \eta^{\nu\eta\tau}\widetilde{X}_{\rho\tau}{Z}_{\eta}~^{\rho}.
\end{equation}
where we defined a new tensor $Z^{\mu\nu}$ as
\begin{equation}
Z^{\mu\nu}=\eta^{\mu\alpha\beta}\nabla_{\alpha}\widetilde{X}_{\beta}~^{\nu}.
\end{equation}
 Let us now check the properties of the $Z$-tensor. It is traceless, but it is not automatically symmetric as can be seen from
\begin{equation}
\eta_{\mu\nu\sigma}Z^{\mu\nu}=-\nabla_{\nu}\widetilde{X}_{\sigma}~^{\nu}+\nabla_{\sigma}\widetilde{X}.
\end{equation}
But with the choice $a=-\frac{1}{2}$ ,  $Z^{\mu\nu}$ becomes symmetric. So we make this choice which yields $\widetilde{X}_{\sigma\eta}=X_{\sigma\eta}-\frac{1}{2}g_{\sigma\eta}X$ and 
$\widetilde{X}=-\frac{1}{2}X$ . With these, one can compute the divergence of $Z_{\mu\nu}$  as
\begin{equation}
\nabla_{\mu}Z^{\mu\nu}=\eta^{\nu\alpha\beta}R_{\alpha\lambda}\widetilde{X}_{\beta}~^{\lambda}=\eta^{\nu\alpha\beta}G_{\alpha\lambda}\widetilde{X}_{\beta}~^{\lambda}=\eta^{\nu\alpha\beta}S_{\alpha\lambda}\widetilde{X}_{\beta}~^{\lambda},
\end{equation}
which is clearly nice as we have started to see the tensors related to the metric, {\it i.e.} the Einstein or the Schouten tensor. The last equation vanishes without the use of any field equation (which we have not yet introduced)  if  $\widetilde{X}_{\beta}~^{\lambda}$ is of the form
\begin{equation}
\widetilde{X}_{\beta}~^{\lambda}=a_{0}\delta_{\beta}~^{\lambda}+a_{1}S_{\beta}~^{\lambda}+a_{2}{{\cal{S}}_2}_{\beta}~^{\lambda}+a_{3}{{\cal{S}}_3}_{\beta}~^{\lambda}+a_{4}{{\cal{S}}_4}_{\beta}~^{\lambda}+ \sum_{i=5}^\infty a_i  {{\cal{S}}_i}_{\beta}~^{\lambda}.
\end{equation}
Note that we do not introduce any derivative terms, as they will bring in extra propagating degrees of
freedom when we build our field equations.  We have separated  the powers beyond 4 as they will not yield 
independent two-tensor structures, due to the Schouten identities, as shown below.  And moreover, for this section, let us stay at the quadratic order and deal with the cubic and quartic order terms in the next section.
So $\widetilde{X}$ reads
\begin{equation}
\widetilde{X}=3a_{0}+a_{1}S+a_{2}{\cal{S}}_2.
\end{equation}
From $X_{\sigma\eta}=\widetilde{X}_{\sigma\eta}-g_{\sigma\eta}\widetilde{X}$, one obtains the ${X}$-tensor as
\begin{equation}
X_{\sigma\eta}=-2g_{\sigma\eta}a_{0}+a_{1}(S_{\sigma\eta}-g_{\sigma\eta}S)+a_{2}\big ( {{\cal S}_2}_{\sigma\eta}-g_{\sigma\eta}{\cal{S}}_2 \big ),
\end{equation}
or in terms of Einstein tensor, one has
\begin{equation}
X_{\sigma\eta}=-2g_{\sigma\eta}a_{0}+a_{1}G_{\sigma\eta}+a_{2} \big (G_{\sigma}\ ^{\mu}G_{\mu\eta}+\frac{R}{2}G_{\sigma\eta}+\frac{R^{2}}{8}g_{\sigma\eta}-g_{\sigma\eta}G_{\mu\nu}^{2} \big ).
\end{equation}
We assumed that the covariant divergence of $X^{\mu \nu}$ vanishes which is possible if and only if $a_2=0$. Then the $Z^{\mu\nu}$ reads 
\begin{equation}
Z^{\mu\nu}=a_{1}C^{\mu\nu},
\end{equation}
which leads to
\begin{equation}
\nabla_{\mu}Y^{\mu\nu}=\,\eta^{\nu\eta\tau}\widetilde{X}_{\rho\tau}{Z}_{\eta}~^{\rho} = a_1 \eta^{\nu\eta\tau} ( a_0 g_{ \rho \tau} + a_1 S_{\rho  \tau} )  C_\eta\,^\rho,
\end{equation}
which vanishes on shell for the field equations 
\begin{equation}
C_{\mu \nu} =  c_1 g_{\mu \nu} + c_2 S_{\mu \nu} + c_3 Y_{\mu \nu},
\end{equation}
which is just MMG   with $ Y^{\mu \nu} = J^{\mu \nu}$ proving the uniqueness of the theory at the quadratic order. Let us now move on to the cubic and quartic powers. 

\section{$R^{3}$ and $R^{4 }$ extensions ?}

\subsection{  $R^{3}$ extension}

Suppose we have the following deformation of TMG and MMG,
\begin{equation}
\sigma G_{\mu\nu}+\Lambda g_{\mu\nu}+\frac{1}{\mu}C_{\mu\nu}+\gamma_1 J_{\mu\nu} + \gamma_2 {\cal{ K}}_{\mu\nu}=0,
\label{fieldeqn3}
\end{equation}
with the most general two-tensor ${\cal{ K}}_{\mu\nu}$  built from the powers of the Ricci tensor and not from its derivatives. (Needless to say, since the Ricci and Riemann tensors are double duals of each other in three dimensions, one does not consider the Riemann tensor.) Therefore, one has  the following tensor: 
\begin{equation}
\begin{split}
{\cal{ K}}^{\mu\nu}\equiv a_{1} {\cal{R}}_3^{\mu\nu} +a_{2}g^{\mu\nu} {\cal{R}}_3
+a_{3}R {\cal{R}}_2^{\mu\nu} +a_{4}R^{\mu\nu} {\cal{R}}_2+a_{5}g^{\mu\nu}R{\cal{R}}_2
+a_{6}R^{\mu\nu}R^{2}+a_{7}g^{\mu\nu}R^{3}.
\end{split}
\end{equation}
We should note that one can eliminate one of the terms since not all of these, ostensibly, algebraically independent terms are actually independent. The quickest way to see this is to use the Cayley-Hamilton  theorem.  At the end of this discussion, we shall make use of this theorem, but for now,  let us proceed with this form of the ${\cal{ K}}$-tensor. Its trace is
reads 
\begin{equation}
\begin{split}
{\cal{ K}}=  ( a_{1}+ 3 a_2){\cal{R}}_3
+(  a_3 + a_{4} + 3 a_5) R{\cal{R}}_2
+(a_{6}+ 3 a_7) R^{3} ,
\end{split}
\end{equation}
and its  covariant-divergence can be computed as  
\begin{equation}
\begin{split}
\nabla_{\mu}{\cal{ K}}^{\mu\nu}=
\nabla_{\mu}R \Bigg ( (a_{1}+a_{3}) {\cal{R}}_2^{\mu\nu}+ \Big (\frac{3a_{3}}{4}+2a_{6}+\frac{a_{4}}{2}\Big)RR^{\mu\nu} 
+\Big (\frac{a_{4}}{2}+a_{5}+\frac{3a_{2}}{4} \Big  ) g_{\mu \nu} {\cal{R}}_2\\+\Big(\frac{a_{5}}{2}+\frac{a_{6}}{2}+3a_{7}\Big)  g_{\mu \nu} R^{2} \Bigg ) 
+{\cal{R}}_2^{\alpha\rho} \Big(3a_{2}\nabla^{\nu}S_{\rho\alpha}+a_{1}\nabla_{\alpha}S_\rho\,^\nu \Big) +R^{\mu\nu}R_{\alpha\beta} \Big (a_{1}\nabla_{\alpha}S_{\beta\mu}+2a_{4}\nabla_{\mu}S_{\beta\alpha}\Big) \\
+RR^{\mu\rho}\Big (a_{3}\nabla_{\mu}S_{\rho}^{\nu}+2a_{5}\nabla^{\nu}S_{\mu\rho}\Big).
\end{split}
\end{equation}
For this divergence to vanish on shell of the theory (\ref{fieldeqn3}), one must see the appearance of the Cotton, Einstein or the $J$-tensors. For  this purpose, one should turn the last three terms into the Cotton tensors and the terms multiplying $\nabla_\mu R$  to the $J$-tensor. These, respectively, can be achieved  if one sets
\begin{equation}
a_1 + 2 a_4=0, \hskip  0,8 cm  a_1 + 3a_2 =0, \hskip 0,8 cm a_3 + 2a_5 =0,
\end{equation}
and
\begin{equation}
\begin{split}
a_{1}+a_{3}= k,  \hskip 0.4 cm a_{3}+  \frac{8}{3} a_6 + \frac{2}{3} a_4 =- k, \hskip 0.3 cm   a_{4}+2a_{5} + \frac{3}{2} a_2=-k  \hskip 0.3 cm 
a_{6}+a_{5}+6a_{7}= \frac{5}{8}k .
\end{split}
\end{equation}
These reduce the divergence of the ${\cal{ K}}^{\mu\nu}$-tensor to 
\begin{equation}
\begin{split}\nabla_{\mu}{\cal{ K}}^{\mu\nu}=  k J^{\mu \nu} \nabla_{\mu}R  + a_1 \eta_{\lambda}~^{\nu\alpha}R_{\alpha\beta}R^{\beta}~_{\rho}C^{\lambda\rho}+a_1 \eta_{\lambda\mu\alpha}R^{\mu\nu}R_{\alpha\beta}C^{\lambda\beta}+ a_3 \eta_{\lambda\mu}~^{\nu}RR^{\mu\rho}C^{\lambda}~_{\rho},
\label{div-K}
\end{split}
\end{equation} 
where we have also made use of (\ref{3indexcot}) and (\ref{2indexcot}).  The penultimate term vanishes due to symmetries,  and one can combine the remaining two terms that have Cotton tensor using the three-dimensional identity  valid for any vector $\xi_\mu$,
\begin{equation}
\eta^{\lambda\nu\alpha}\xi^{\rho}=g^{\lambda\rho}\eta^{\beta\nu\alpha}\xi_{\beta}+g^{\nu\rho}\eta^{\lambda\beta\alpha}\xi_{\beta}+g^{\alpha\rho}\eta^{\lambda\nu\beta}\xi_{\beta},
\end{equation}
as
\begin{equation}
\eta_{\lambda}\thinspace^{\nu\alpha}R^{\beta}\thinspace_{\rho}=\delta_{\lambda}^{\beta}\eta^{\sigma\nu\alpha}R_{\sigma\rho}
+g^{\nu\beta}\eta_{\lambda}\thinspace^{\sigma\alpha}R_{\sigma\rho}+g^{\alpha\beta}\eta_{\lambda}\thinspace^{\nu\sigma}R_{\sigma\rho},
\end{equation}
to arrive at
\begin{equation}
\begin{split}\nabla_{\mu}{\cal{ K}}^{\mu\nu}=  k \Bigg ( J^{\mu \nu} \nabla_{\mu}R  +  \eta_{\lambda\mu}~^{\nu}RR^{\mu\rho}C^{\lambda}~_{\rho} \Bigg).
\label{div-K2}
\end{split}
\end{equation}
The trace reduces to a simple expression in terms of the trace of the $J$-tensor as 
\begin{equation}
{\cal{ K}}=  -\frac{k}{2} R \Big ( R_{\alpha\beta}^{2}- \frac{3 }{8}  R^{2} \Big ) =  - k R J.
\label{trace}
\end{equation}
The last two equations are all we need to find the $K$-tensor that could possibly vanish on the  TMG or MMG shell.  It is important to realize that the number $k$ plays a crucial role here. If $k=0$,  then clearly {\it without } using the field equations. ${\cal{ K}}^{\mu\nu}$ is conserved, and it is traceless.  Explicitly one has 
\begin{equation}
\begin{split}
{\cal{ K}}^{\mu\nu}={\cal{R}}_3^{\mu\nu}-\frac{1}{3}g^{\mu\nu}{\cal{R}}_3-R{\cal{R}}_2^{\mu\nu}-\frac{1}{2}R^{\mu\nu}{\cal{R}}_2 +\frac{1}{2}g^{\mu\nu}R{\cal{R}}_2+\frac{1}{2}R^{\mu\nu}R^{2}-\frac{1}{6}g^{\mu\nu}R^{3}.
\end{split}
\end{equation}
But this is a red herring: as the Cayley-Hamilton theorem shows, this tensor is identically zero.
Now consider a $3\times 3$ matrix $A$;  then this matrix satisfies the same equation as its eigenvalues :
\begin{equation}
A^{3}-(\mbox{Tr}A)A^{2}+\frac{1}{2}\left[(\mbox{Tr}A)^{2}-\mbox{Tr}(A^{2})\right]A-\mbox{det}(A)I_{3}=0.
\end{equation}
Taking the trace of this equation, one has the determinant in terms of traces as
\begin{equation}
\mbox{det} A=\frac{1}{6}\left[(\mbox{Tr}A)^{3}-3\mbox{Tr}(A^{2})(\mbox{Tr}A)+2\mbox{Tr}(A^{3})\right].
\end{equation}
These two equations for the matrix  $A = (R^\mu_\nu)$ yield ${\cal{ K}}_{\mu\nu}=0$. The second option is to consider $ k \ne 0$, and  then in (\ref{div-K2}), the second term 
vanishes both on the TMG and MMG mass shell, but the first term does not vanish. One could ask whether the theory, as in TMG, requires  $\nabla_\mu R = 0$, which is not so,as is clear from the trace equation (\ref{trace}).  Hence, there does not exist a nontrivial tensor cubic in the curvature that could be used to deform TMG or MMG while keeping its single particle content intact.

\subsection{  $R^{4}$ extension}

 The most general two-tensor built with the powers of the Ricci tensor is 
\begin{equation}
\begin{split}
{\cal{ L}}^{\mu\nu}=a_{1}{\cal R}_4^{\mu \nu}+a_{2}{\cal R}_2^{\mu \nu}{\cal R}_2 +a_{3}R{\cal R}_3^{\mu \nu}+a_{4}R^{2}{\cal R}_2^{\mu \nu}
+a_{5}R^{\mu\nu}R^{3}+a_{6}R^{\mu\nu}{\cal R}_3\\
+a_{7}R^{\mu\nu}R{\cal R}_2+a_{8}g^{\mu\nu}{\cal R}_4
+a_{9}g^{\mu\nu}R{\cal R}_3+a_{10}g^{\mu\nu}R^2{\cal R}_2+a_{11}g^{\mu\nu}R^{4}+a_{12}g^{\mu\nu}{\cal R}_2^2.
\end{split}
\end{equation}
Due to Schouten identity and the fact that $\cal{K}^{\mu \nu}$ is zero, not all terms are linearly independent in this tensor, but we shall work with this general form and eliminate the dependent  terms later. Then its
divergence follows as 
\begin{equation}
\begin{split}
\nabla_\mu {\cal{ L}}^{\mu\nu}=\Bigg ((\frac{5}{4}a_{1}+a_{3}){\cal R}_3^{\mu \nu}+(\frac{3}{4}a_{2}+\frac{3}{4}a_{6}+a_{7})R^{\mu\nu}{\cal R}_2^{\mu \nu}
+(a_{3}+2a_{4}+\frac{1}{2}a_{2})R{\cal R}_2^{\mu \nu} \\
+(\frac{3}{4}a_{4}+\frac{1}{2}a_{7}+3a_{5})R^{2}R^{\mu\nu}+(\frac{1}{2}a_{5}+4a_{11}+\frac{1}{2}a_{10})g^{\mu\nu}R^{3}+\\(\frac{1}{2}a_{6}+a_{9}+a_{8})g^{\mu\nu}{\cal R}_3
+(\frac{1}{2}a_{7}+a_{12}+2a_{10}+\frac{3}{4}a_{9})g^{\mu\nu}R{\cal R}_2 \Bigg )\nabla_{\mu}R\\
+R^{\mu\alpha}{\cal R}_2(a_{2}\nabla_{\mu}S_{\alpha}~^{\nu}+4a_{12}\nabla^{\nu}S_{\mu\alpha})+R{\cal R}_2^\mu\,_\beta (a_{3}\nabla_{\mu}S^{\beta\nu}+3a_{9}\nabla^{\nu}S_{\mu}~^{\beta})\\
+RR^{\mu\alpha}R^{\beta\nu}(a_{3}\nabla_{\mu}S_{\alpha\beta}+2a_{7}\nabla_{\beta}S_{\alpha\mu})+R^{2}R^{\mu\alpha}(a_{4}\nabla_{\mu}S_{\alpha}~^{\nu}+2a_{10}\nabla^{\nu}S_{\mu\alpha})\\+{\cal R}_2^\mu\,_\beta R_{\rho}~^{\nu}(a_{1}\nabla_{\mu}S^{\beta\rho}+3a_{6}\nabla^{\rho}S_{\mu}~^{\beta})
+{\cal R}_3^{\mu \rho}(a_{1}\nabla_{\mu}S_{\rho}~^{\nu}+4a_{8}\nabla^{\nu}S_{\mu\rho})\\+R^{\mu\alpha}{\cal R}_2^{\nu \beta} (a_{1}\nabla_{\mu}S_{\alpha\beta}+2a_{2}\nabla_{\beta}S_{\alpha\mu}).
\end{split}
\end{equation}
The terms in the last four lines can be written in terms of the Cotton tensor only if  the numerical parameters are related as 
\begin{equation}
a_{2}=-\frac{a_{1}}{2},\,\,\,\,a_{6}=-\frac{a_{1}}{3},\,\,\,\, a_{7}=-\frac{a_{3}}{2},~a_{8}=-\frac{a_{1}}{4},~ a_{9}=-\frac{a_{3}}{3},~ a_{10}=-\frac{a_{4}}{2},~a_{12}=\frac{a_{1}}{8}.
\end{equation}
The terms multiplying the derivative of the curvature scalar gives rise to the $J$-tensor when the parameters are tuned as 
\begin{equation}
\frac{5a_{1}}{4}+a_{3}=- k  \hskip 0.3 cm a_{3}-  \frac{1}{4} a_1 +2 a_4 = \frac{3k}{4}, \hskip 0.3 cm   -  \frac{1}{4} a_3+3a_{5} + \frac{3}{4} a_4=-\frac{5k}{16} , \hskip 0.3 cm 
-\frac{1}{4}a_{4}+\frac{1}{2}a_{5}+4a_{11}= \frac{17}{192}k .
\end{equation}
This linear equation set is solved for all $a_{i}$ in terms of $a_1$ and $k$, upon use of which one arrives at  the divergence as 
\begin{equation}
\begin{split} 
\nabla_\mu {\cal L}^{\mu\nu}=k\Bigg (R^{\mu\alpha}J_{\alpha}~^{\nu}-\frac{1}{3}g^{\mu\nu}(R^{\alpha\beta}J_{\alpha\beta}-\frac{1}{8}RJ)\Bigg )\nabla_{\mu}R+\Bigg (\frac{1}{2}a_{1} { \cal R}_2+(\frac{k}{8}+\frac{a_{1}}{2})R^{2}\Bigg )\nabla_{\mu}J^{\mu\nu}\\-a_{1}RR^{\mu\nu}\nabla_{\alpha}J^{\alpha}_{\mu}+a_1\eta_{k}~^{\nu}~_{\mu}{\cal R}_3^{\mu\rho} C^{k}~_{\rho},
\label{div-L}
\end{split}
\end{equation}
and  the trace as
\begin{equation}
{\cal L}=\frac{k}{8}R^{2}J.
\end{equation}
From the trace, we learn that in general the curvature scalar will not be constant since the $J$-tensor has the square of the Ricci tensor in it, and hence we must set  $k=0$ for the first term in the divergence to vanish since the term in the parentheses is not generically zero.   The other terms in (\ref{div-L}) vanish on shell.  Once again, we seem to have gotten an on shell-conserved tensor, but it turns out that this tensor given as 
\begin{equation}
\begin{split}
{\cal{ L}}^{\mu\nu}={\cal R}_4^{\mu \nu} -\frac{1}{2}{\cal R}_2^{\mu \nu}{\cal R}_2 - \frac{5}{4}R{\cal R}_3^{\mu \nu}+ \frac{3}{4}R^{2}{\cal R}_2^{\mu \nu}
-\frac{7}{24}R^{\mu\nu}R^{3}- \frac{1}{3}R^{\mu\nu}{\cal R}_3\\
+\frac{5}{8}R^{\mu\nu}R{\cal R}_2-\frac{1}{4}g^{\mu\nu}{\cal R}_4
+\frac{5}{12}g^{\mu\nu}R{\cal R}_3-\frac{3}{8}g^{\mu\nu}R^2{\cal R}_2-\frac{1}{12}g^{\mu\nu}R^{4}+\frac{1}{8}g^{\mu\nu}{\cal R}_2^2.
\end{split}
\end{equation}
 is identically zero, if one uses the fact that 
$R^{\mu  \rho } {\cal K}^\mu\,_\rho =0$ which  yields 
\begin{equation}
\begin{split}
{\cal R}_4^{\mu \nu }= \frac{1}{3}R^{\mu\nu}{\cal{R}}_3+R{\cal{R}}_3^{\mu\nu}+\frac{1}{2}{\cal R}_2^{\mu\nu}{\cal{R}}_2 -\frac{1}{2}R^{\mu\nu}R{\cal{R}}_2-\frac{1}{2}{\cal R}_2^{\mu\nu}R^{2}+\frac{1}{6}R^{\mu\nu}R^{3},
\end{split}
\end{equation}
and its trace 
\begin{equation}
{\cal R}_4=\frac{4}{3}R{\cal R}_3+\frac{1}{2}{\cal{R}}_2^2-R^{2}{\cal{R}}_2+\frac{1}{6}R^{4}.
\label{tracer4}
\end{equation}
Since ${\cal L}^{\mu\nu}=0$  identically, there are no nontrivial quartic extensions of TMG and MMG. Beyond the  quartic order, it is easy to show that all the possible rank-2 tensors built form the powers of the curvature can be written in terms of the lower order ones \cite{gurses}.
To see this, let us denote the traceless Ricci tensor as  $  \tilde{R}_{\mu\nu}$; then one has 
\begin{equation}
\delta_{[\mu_{1}\mu_{2}\mu_{3}\mu_{4}]}^{\nu_1\nu_{2}\nu_3\nu_{4}}\widetilde{R}_{\nu_{1}}^{\mu_{2}}\widetilde{R}_{\nu_{2}}^{\mu_{3}}\widetilde{R}_{\nu_{3}}^{\mu_{4}}\widetilde{R}_{\nu_{4}}^{\mu_{1}}=\frac{1}{4}\widetilde{\cal{R}}_{4}-\frac{1}{8}\widetilde{\cal{R}}_{2}^2=0,
\end{equation}
where the bracket represents the total antisymmetrization. The result  is just the same as (\ref{tracer4}) written in the traceless tensors. The more important object is the rank-2 tensor 
\begin{equation}
\delta_{[\mu_{1}\mu_{2}\mu_{3}\mu_{4}]}^{\nu_1\nu_{2}\nu_3\nu_{4}}\widetilde{R}_{\nu_{1}}^{\mu_{2}}\widetilde{R}_{\nu_{2}}^{\mu_{3}}\widetilde{R}_{\nu_{3}}^{\mu_{4}}\widetilde{R}_{\nu_{4}}^{\mu}\widetilde{R}_{\nu}^{\mu_{1}}=\frac{1}{4} ({\widetilde{\cal{R}}_{5}})_\nu^\mu-\frac{1}{8} \widetilde{\cal{R}}_{2} ({\widetilde{\cal{R}}_{3}})_\nu^\mu
-\frac{1}{12}\widetilde{\cal{R}}_{3} ({\widetilde{\cal{R}}_{2}})_\nu^\mu=0,
\end{equation}
which proves the claim. Therefore, there does not exist a nontirvial  on TMG-shell conserved rank-2 tensor beyond the quadratic one already found in \cite{Townsend1}. 

\section{Conclusions}

In this work, we have made an exhaustive search of possible deformations of the topologically massive gravity beyond the minimal massive gravity, with the condition that the single massive degree of freedom is intact, and have shown that no such deformations exist. Minimal massive gravity is a rather unique theory improving 
the boundary behavior of TMG while keeping its bulk properties intact. Therefore it is a candidate model which might have a dual unitary boundary conformal field theory unlike the other three-dimensional gravity theories. The model has been subject to recent works  both in terms of classical solutions and  in terms of semiclassical analysis besides the ones we quoted before in \cite{Arv2, Altas,setare,gaston,Yekta,deger,Ali}. With this work, we have also shown that it is highly difficult to construct on-shell conserved rank-2 tensors in three dimensions, a question which needs to be studied in higher dimensions.  It would also be of some interest to extend these models to the ones with two massive degrees of freedom, extending the work initiated in \cite{mmg2}.

 The work of B.T. was supported by TUBITAK Grant No. 113F155.

\end{document}